\documentclass[10pt,english,prl,floatfix,superscriptaddress,twocolumn,showpacs,amsmath,amssymb,reprint]{revtex4-1}
\usepackage[T1]{fontenc}
\usepackage[latin9]{inputenc}
\usepackage{verbatim}
\usepackage{amsmath}
\usepackage{amssymb}

\makeatletter
\PassOptionsToPackage{caption=false}{subfig} 
\usepackage{hyperref}
\hypersetup{
breaklinks=true,
colorlinks=true,
citecolor=blue,
linkcolor=blue,
filecolor=blue,
urlcolor=blue
}
\IfFileExists{lmodern.sty}{\usepackage{lmodern}}{}
\makeatother

\usepackage{babel}
\begin{document}
\title{A class of solutions of the two-dimensional Toda lattice equation}
\author{V. N. Duarte}
\email{vduarte@pppl.gov}

\address{Princeton Plasma Physics Laboratory, Princeton University, Princeton,
NJ, 08543, USA}
\date{\today}
\begin{abstract}
A method is proposed to systematically generate solutions of the two-dimenional
Toda lattice equation in terms of previously known solutions $\phi\left(x,y\right)$
of the two-dimensional Laplace's equation. The two-dimensional solution
of Nakamura's {[}J. Phys. Soc. Jpn. \textbf{52}, 380 (1983){]} is
shown to correspond to one particular choice of $\phi\left(x,y\right)$.
\end{abstract}
\maketitle

The numerical discovery of solitons in collisionless plasmas \citep{Zabusky},
accompanied by an explanation for the recurrence of states in nonlinear
prototype string systems \citep{fermi1955studies}, and the subsequent
development of the inverse scattering tranform to integrate the Korteweg-de
Vries equation \citep{GardnerGreeneKruskalMiura,Lax1968} led to an
intense search for exactly solvable, completely integrable systems
described by dispersive nonlinear partial differential equations.
More modern applications of such systems include topological solitons
in the context of magnetic skyrmions \citep{bogdanov2020physical}
and the Wess\textendash Zumino\textendash Witten model \citep{WESS197195,WITTEN1983422}.

Among the mentioned class of exactly solvable systems is Toda's devise
of a potential form that couples nearest neighbors in a lattice that
allows for complete integration of the equation that governs the vibrations
of the lattice structure \citep{toda1967vibration}. In the continuum
limit, the one-dimensional Toda equation recovers the Korteweg-de
Vries equation \citep{toda2012theory} while the two-dimensional Toda
equation recovers \citep{Hirota1981Kadomtsev} the Kadomtsev-Petviashvili
equation, originally derived to model the effect of long transverse
perturbations on the dynamics of plasma ion-acoustic modes of long
wavelength and small amplitude \citep{kadomtsev1970stability}.

For many nonlinear problems, it is useful to have their solutions
expressed in terms of solutions to linear problems, as is the case
of approaches that express the solutions of Liouville's equation in
terms of solutions of Laplace's equation \citep{walker1915some,ClementeLiouville1992}
and the case of the Cole-Hopf transform \citep{cole1951quasi,hopf1950partial}
that maps solutions of the Burgers' equation onto solutions of a linear
diffusion equation. Inspired by the former, this note develops a technique
to express solutions, both solitonic and non-solitonic, of the two-dimensional
Toda lattice equation (with two continuous variables $X,Y$ and one
discrete variable $Z\equiv n$) in terms of solutions of Laplace's
equation, for which existence and uniqueness of solutions are guaranteed
for given boundary conditions. The two-dimensional Toda lattice equation
with its corresponding exponential restoring force reads \citep{nakamura1983exact,toda2012theory}

\begin{multline}
\alpha u_{XX}\left(X,Y,n\right)+\beta u_{YY}\left(X,Y,n\right)=\\
=e^{u\left(X,Y,n-1\right)-u\left(X,Y,n\right)}-e^{u\left(X,Y,n\right)-u\left(X,Y,n+1\right)}.\label{eq:ZerothEq}
\end{multline}
To exploit the properties of Laplace's equation, the continuous variables
will be rescaled as 
\begin{equation}
x\equiv X/\sqrt{\alpha},\:\:y\equiv Y/\sqrt{\beta},\label{eq:Normalization}
\end{equation}
hence

\begin{multline}
u_{xx}\left(x,y,n\right)+u_{yy}\left(x,y,n\right)=\\
=e^{u\left(x,y,n-1\right)-u\left(x,y,n\right)}-e^{u\left(x,y,n\right)-u\left(x,y,n+1\right)}.\label{eq:FirstEq}
\end{multline}
Solutions to Eq. \eqref{eq:FirstEq} will be sought using the ansatz
\begin{equation}
u\left(x,y,n\right)=F\left(\phi\left(x,y\right),n\right)+\ln\left[\left|\nabla\phi\left(x,y\right)\right|^{-2n}\right],\label{eq:ansatz}
\end{equation}
where $\phi\left(x,y\right)$ is a solution of the two-dimensional
Laplace's equation
\begin{equation}
\phi_{xx}+\phi_{yy}=0\label{eq:LaplaceEq}
\end{equation}
 and $F\left(\phi,n\right)$ is a function to be determined. The present
derivation takes advantage of the property noted by Clemente \citep{ClementeLiouville1992}
that, if \eqref{eq:LaplaceEq} holds, then also 
\begin{equation}
\left[\ln\left(\left|\nabla\phi\right|^{\gamma}\right)\right]_{xx}+\left[\ln\left(\left|\nabla\phi\right|^{\gamma}\right)\right]_{yy}=0\label{eq:LaplaceEqforLog}
\end{equation}
is true $\forall\gamma\in\mathbb{C}$. In addition, the calculation
will use the fact that, for any function $\psi\equiv\psi\left(\phi\left(x,y\right),n\right)$,
the relation \footnote{This property has been extensively used in tokamak plasma modeling,
to formally construct solutions for the poloidal flux stream function
$\psi$ of a pressure-anisotropic axisymmetric equilibrium in terms
of solutions of the isotropic Grad-Shafranov equation \citep{clemente1993anisotropic}.
Such technique involves an integral transform and can also be employed
in the presence of plasma rotation \citep{Clemente_1994}.}
\begin{equation}
\nabla^{2}\psi=\frac{\partial\psi}{\partial\phi}\nabla^{2}\phi+\frac{\partial^{2}\psi}{\partial\phi^{2}}\left|\nabla\phi\right|^{2}\label{eq:ChainRule}
\end{equation}
holds as long as the Laplacian and the gradient are two-demensional
in $x,y$ and therefore independent of the variable $n$. Subtituting
the ansatz \eqref{eq:ansatz} into Eq. \eqref{eq:FirstEq} and simplifying
the resulting expression using Eqs. \eqref{eq:LaplaceEq}, \eqref{eq:LaplaceEqforLog}
and \eqref{eq:ChainRule}, the dimensionality of the problem is reduced
from $\left(x,y,n\right)$ to $\left(\phi,n\right)$ and Eq. \eqref{eq:FirstEq}
becomes

\begin{multline}
\frac{\partial^{2}}{\partial\phi^{2}}F\left(\phi\left(x,y\right),n\right)=e^{F\left(\phi\left(x,y\right),n-1\right)-F\left(\phi\left(x,y\right),n\right)}-\\
-e^{F\left(\phi\left(x,y\right),n\right)-F\left(\phi\left(x,y\right),n+1\right)},\label{eq:F}
\end{multline}
Note that the ansatz \eqref{eq:ansatz} was chosen in such a way as
to exactly cancel out the dependencies on $\left|\nabla\phi\right|$
that otherwise would be present in Eq. \eqref{eq:F}. The fact that
formally Eq. \eqref{eq:F} only depends on two independent variables,
rather than three as is the case of Eqs. \eqref{eq:ZerothEq} and
\eqref{eq:FirstEq}, allows for a simplified construction of conservation
laws relative to Ref. \citep{KajiwaraToda1991}. Eq. \eqref{eq:F}
can be expressed in terms of the function $r\left(\phi,n\right)\equiv F\left(\phi,n\right)-F\left(\phi,n-1\right)$
as
\begin{equation}
\frac{\partial^{2}r\left(\phi,n\right)}{\partial\phi^{2}}=2e^{r\left(\phi,n\right)}-e^{r\left(\phi,n-1\right)}-e^{r\left(\phi,n+1\right)}\label{eq:EqForr}
\end{equation}
Defining $f\left(\phi,n\right)$ via
\begin{equation}
e^{-F\left(\phi,n\right)+F\left(\phi,n-1\right)}=e^{-r\left(\phi,n\right)}=c+\frac{\partial^{2}}{\partial\phi^{2}}\ln f\left(\phi,n\right),\label{eq:ChoiceForf}
\end{equation}
with $c$ being a constant, allows Eq. \eqref{eq:EqForr} to be cast
in Hirota's bilinear form \citep{hirota1974new}, following a similar
procedure as done in the one-dimensional case \citep{toda2012theory}.
The substitution of Eq. \eqref{eq:ChoiceForf} into Eq. \eqref{eq:EqForr}
leads to 
\begin{equation}
e^{-F\left(\phi,n\right)+F\left(\phi,n-1\right)}=\frac{f\left(\phi,n+1\right)f\left(\phi,n-1\right)}{f^{2}\left(\phi,n\right)}e^{c_{1}\phi+c_{2}}\label{eq:FinalNakamura}
\end{equation}
where $c_{1}$ and $c_{2}$ are constants of integration. A solution
can then be obtained through the ansatz 
\begin{equation}
f\left(\phi,n\right)=1+e^{p\phi+\bar{\omega}n},\label{eq:ansatzForf}
\end{equation}
which yields
\begin{equation}
e^{-F\left(\phi,n\right)+F\left(\phi,n-1\right)}=\left[1+\frac{p^{2}}{4}sech^{2}\left(\frac{p\phi+\bar{\omega}n}{2}\right)\right]e^{c_{1}\phi+c_{2}},\label{eq:SolForF}
\end{equation}
where $\sinh\left(\bar{\omega}/2\right)=\pm p/2$. Or, in terms of
the original function as defined by \eqref{eq:ansatz},

\begin{multline}
e^{-u\left(x,y,n\right)+u\left(x,y,n-1\right)}=\\
=\left[1+\frac{p^{2}}{4}sech^{2}\left(\frac{p\phi+\bar{\omega}n}{2}\right)\right]\frac{e^{c_{1}\phi+c_{2}}}{\left|\nabla\phi\right|^{2}}.\label{eq:SolForu}
\end{multline}

The one-soliton solution of Ref. \citep{nakamura1983exact} is exactly
recovered with the particular choice of a traveling wave form for
the Laplace's equation solution $\phi$, linear in both $x$ and $y$,
specifically $\phi\left(x,y\right)=\left(\sqrt{\alpha}kx+\sqrt{\beta}ly\right)/p$
with $p=\pm\sqrt{\alpha k^{2}+\beta l^{2}}$, $c_{1}=c_{2}=0$. In
that case $\left|\nabla\phi\left(x,y\right)\right|^{2}=1$. Using
these choices and returning to the orginal variables using Eq. \eqref{eq:Normalization},
Eq. \eqref{eq:SolForu} becomes

\begin{multline}
e^{-u\left(X,Y,n\right)+u\left(X,Y,n-1\right)}-1=\\
=\frac{\alpha k^{2}+\beta l^{2}}{4}sech^{2}\left[\left(kX+lY+\bar{\omega}n\right)/2\right].\label{eq:SolForuNakamura}
\end{multline}
Eq. \eqref{eq:SolForuNakamura} is the same as Eq. 3.8 of Nakamura's
paper \citep{nakamura1983exact}. It can be appreciated that the most
stringent assumption of the present derivation was to assume a form
for $f$ (Eq. \eqref{eq:ansatzForf}). It should be noted, however,
that other solutions, more convenient for a given physical problem
at hand, can be constructed for different forms of $f$ but still
using the same methodology presented in this work, i.e., different
$f$ will lead to distinct $F\left(\phi,n\right)$ while still preserving
the same solution ansatz (Eq. \eqref{eq:ansatz}). Because of its
$e^{c_{1}\phi}$ ($c_{1}\in\mathbb{C}$) dependence, Eq. \eqref{eq:SolForu}
provides a direct means for constructing breather-like solutions \citep{flytzanis1985kink},
i.e., solutions with localization due to amplitude decay in the continuous
variables and oscillation in the discrete variable or vice-versa,
depending on the choice of the constants $c_{1},p$ and $\bar{\omega}$.

In summary, using the serendipitous identity \eqref{eq:LaplaceEqforLog}
and an ansatz \eqref{eq:ansatz} that takes advantage of the structure
of the underlying equation \eqref{eq:FirstEq}, new solutions of the
two-dimensional (nonlinear) Toda lattice equation were constructed
in terms of solutions $\phi$ of the (linear) Laplace's equation.
A particular choice of $\phi$ was shown to replicate a previously
found solution for the lattice equation \citep{nakamura1983exact}.
Other solutions can then be constructed from \eqref{eq:SolForu} to
the limits of  the imposed boundary conditions of a particular problem.
For example, expression \eqref{eq:SolForu} provides a rapid means
for construction of solutions with azimuthal symmetry. In that case,
$\left(1/r\right)\left(r\phi_{r}\right)_{r}=0$, where $r=\sqrt{X^{2}/\alpha+Y^{2}/\beta}$
and the sub-index denotes differentiation. Therefore any $\phi\propto\ln r$
in Eq. \eqref{eq:SolForu} will lead to an azimuthally invariant solution.
\begin{acknowledgments}
Discussions with V. L. Quito are appreciated. This work was supported
by the US Department of Energy under contract DE-AC02-09CH11466.

\begin{comment}

\subsection*{DATA AVAILABILITY}

The data that support the findings of this study are available within
this article.
\end{comment}
\end{acknowledgments}

\bibliographystyle{apsrev4-1}
\addcontentsline{toc}{section}{\refname}

\end{document}